\documentclass[prd,aps,preprint,amsmath,nofootinbib,amssymb,eqsecnum,showkeys,tightenlines]{revtex4-1}
\usepackage{slashed}
\usepackage{epsfig,latexsym,cancel,amssymb,amsmath,verbatim,mathrsfs}
\usepackage{color}
\usepackage{graphicx}

\def\ra{\rightarrow}
\def\L{\left(}
\def\R{\right)}

\def\f{\frac}
\newcommand{\be}{\begin{equation}}
\newcommand{\ee}{\end{equation}}
\newcommand{\bea}{\begin{eqnarray}}
\newcommand{\eea}{\end{eqnarray}}
\newcommand{\ba}{\begin{array}}
\newcommand{\ea}{\end{array}}

\newcommand{\slg}[1]{#1\hspace{-0.45em{/}}}

\long\def\symbolfootnote[#1]#2{\begingroup%
\def\thefootnote{\fnsymbol{footnote}}\footnote[#1]{#2}\endgroup}

\newcommand{\beq}{\begin{equation}}
\newcommand{\eeq}{\end{equation}}
\begin{document}

\title{$(g-2)_{\mu}$ Versus $K\ra \pi+E_{miss}$ Induced by the $(B-L)_{23}$ Boson}

\author{Zhaofeng Kang}
\email[E-mail: ]{zhaofengkang@gmail.com}
\affiliation{School of Physics, Huazhong University of Science and Technology, Wuhan 430074, China}
%\affiliation{School of Physics, Korea Institute for Advanced Study, Seoul 130-722, Korea}

\author{Yoshihiro Shigekami}
\email[E-mail: ]{sigekami@eken.phys.nagoya-u.ac.jp}
\affiliation{School of Physics, Huazhong University of Science and Technology, Wuhan 430074, China}

\date{\today}

\begin{abstract}

To address the long-standing $(g-2)_{\mu}$ anomaly via a light boson, in Ref.~\cite{Kang:2019vng} we proposed to extend the standard model (SM) by the local $(B-L)_{23}$, under which only the second and third generations of fermions are charged. It predicts an invisible $Z'$ with mass ${\cal O}(100)$ MeV, and moreover it has flavor-changing neutral current (FCNC) couplings to the up-type quarks at tree level. Such a $Z'$, via $K_L \to \pi^0 + Z'(\to \nu \bar{\nu})$ at loop level, may be a natural candidate to account for the recent KOTO anomaly. In this article, we investigate this possibility, to find that $Z'$ can readily do this job if it is no longer responsible for the $(g-2)_{\mu}$ anomaly. We further find that both anomalies can be explained with moderate tuning of the CP violation, but may contradict the $B$ meson decays.

\end{abstract}

\pacs{12.60.Jv, 14.70.Pw, 95.35.+d}

\maketitle

\section{Introduction and experiment reviews}

A dark world far below the weak scale is introduced in many different contexts of new physics beyond the Standard Model (SM). Whether violating the flavor structure of the SM or not, members of the light dark world may imprint in the rare decays of $K$ and $B$ mesons, etc. For instance, it is known many years ago that, a light dark photon which does not have tree-level flavor-changing neutral current (FCNC) couplings to quarks can lead to flavor violation decay $K\to \pi Z'$~\cite{Pospelov:2008zw,Davoudiasl:2012ag}. Hunting hints of such a world is the target of many experiments like BaBar, Belle and LHCb, etc. 

One of the strong motivation for a light dark world is to explain the long-standing $(g-2)_\mu$ puzzle \cite{Terazawa:1968jh,Terazawa:1968mx,Terazawa:1969ih,Hagiwara:2011af,Keshavarzi:2018mgv,Terazawa:2018pdc,Aoyama:2020ynm,Bennett:2006fi,Roberts:2010cj,Davier:2010nc,Davier:2017zfy,Davier:2019can}. To that end, we proposed an extension to the SM by the flavored gauge group $(B-L)_{23}$~\footnote{In the previous paper~\cite{Kang:2019vng}, we call this $(B-L)_{\mu \tau}$. However, more appropriately, we just change the name as $(B-L)_{23}$ in this paper.}, under which only the second and third generations of fermions are charged~\cite{Kang:2019vng}. Therefore, the model furnishes an electron/nucleon phobic muon force $Z'$ with mass $\sim 100$ MeV and gauge coupling $\sim 10^{-3}$, which is capable of explaining the $(g-2)_\mu$ puzzle agreeing with the strong experimental bounds related to the electron and proton. This solution assembles the one using a massive gauge boson~\cite{LmuLtau:g-2,Altmannshofer:2016brv} in the popular gauged $L_\mu-L_\tau$ model~\cite{LmuLtau1,LmuLtau2}.

Largely speaking, the leptonic faces of the $Z'$ in the two models may share common features. However, their difference is obvious when studying the phenomenology of $Z'$ associated with quarks: The $Z'$ from the gauged $L_\mu-L_\tau$ model has no direct couplings to the quarks, whereas our $Z'$ has tree-level FCNC couplings to quarks. In our model building to realize the Cabibbo-Kobayashi-Maskawa (CKM) matrix, we are forced to introduce up-quark-like (vector-like) heavy quarks to connect the first and other two families of up-type quarks, without giving rise to tree-level FCNC couplings between $Z'$ and down-type quarks. Otherwise, rare decays of $K$ and $B$ mesons would place very strong constraints and negates the possibility to explain the $(g-2)_\mu$ discrepancy by that $Z'$. We then claim that the leading flavor violation signature of $Z'$ is from the rare top quark decay $t\ra qZ'(\ra\nu\bar\nu)$, which may have branching ratio $\sim {\cal O}(10^{-4})$, testable at the future colliders~\cite{Kang:2019vng}. However, it may be not true considering that FCNCs in the down-type quark sector can also be induced with the help of a $W$-loop. 

Now it is a good time to study these FCNCs, since recently the KOTO collaborators reported anomalously large signature for $K_L\ra \pi^0 +E_{miss}$, which may hint an invisible light particle with FCNC couplings to the down-type quark sector. We will find that $Z'$ of the local $(B-L)_{23}$ is a very natural candidate to account for the events, but needs a gauge coupling way smaller than the one required to account for the $(g-2)_\mu$ discrepancy. The reason is that the induced FCNCs via the $W$-loop are too large, by virtue of the absence of Glashow-Iliopoulos-Maiani (GIM) suppression. This fact warns us that the original motivation for the local $(B-L)_{23}$ probably fails. Then, we attempt to save it by reducing BR$(K_L\ra \pi^0 +Z')$, focusing on the loophole region of $m_{Z'}$ (around the neutral pion mass region), to which the E949/NA62 search for $K^+\ra \pi^++E_{miss}$ is blind. We find that indeed this is possible at the expense of moderate fine-tuning, thus making the simultaneous explanation to the  $(g-2)_\mu$ and KOTO anomalies feasible. Although this connection between them has already been studied by several groups~\cite{KOTO:2body1,Jho:2020jsa,Liu:2020ser,Dutta:2020scq,Borah:2020swo}, our model may provide the most attractive way to realize the connection: Starting from the gauge symmetry $(B-L)_{23}$, all the ingredients to explain the two anomalies are built-in. The real challenge to our way is how to eliminate the strong tension with the constraints from the rare $B$ meson decays.

The paper is organized as the following: We first give a brief review of the experimental status for $K \to \pi + E_{miss}$ decays in the Section~\ref{sec:exp}. In the Section~\ref{sec:rev}, we explain the profile of $Z'$-induced FCNC for KOTO and introduce our model. In the next section, we calculate the $s \to d$ transition from FCNCs induced by $Z'$ in up-type quark sector. The results of constraints from $K \to \pi Z'$ processes are also shown. In the Section~\ref{sec:Bphys}, we comment on the predictions in $B$ meson decay processes. Section~\ref{sec:conclusion} is devoted to the conclusion.

\section{Searches at KOTO and E949/NA62}
\label{sec:exp}

The KOTO experiment at J-PARC is searching for signature from $K_L\to\pi^0 (2\gamma)+E_{miss}$~\cite{KOTO}, aiming at reaching the SM level BR$(K_L\to\pi^0+\nu\bar \nu)|_{\rm SM}\simeq \L3.4\pm 0.6\R\times 10^{-11}$~\cite{SMBG-1,SMBG-2,SMBG-3}, and now reaches the level $\sim{\cal O}(10^{-9})$~\cite{KOTO}. In the KOTO signal region where the transverse momentum of the reconstructed $\pi^0$ is within the region $130~{\rm MeV}< p_T^{\pi^0}< 250~{\rm MeV}$, the SM expectation merely gives $0.10\pm 0.02$ events. Recently, three events which are distinguishable to the known backgrounds are found \cite{ShionoharaTalk}. Explained by $K_L \to \pi^0 + \nu \bar{\nu}$, it requires an enhancement of the SM branching ratio about two orders of magnitude, 
\begin{equation}
{\rm BR}(K_L\to\pi^0 \nu\bar \nu)|_{\rm KOTO}=2.1^{+2.0(+4.1)}_{-1.1(-1.7)}\times 10^{-9},
\end{equation}
where the uncertainties are due to statistics. However, BR$(K^+ \to \pi^+ + \nu \bar{\nu})$ is also enhanced and then severely constrained by the searches at the E949~\cite{E949-1,E949-2} and NA62~\cite{NA,CortinaGil:2020vlo} experiments. Currently, they set the upper bound BR$(K^+ \to \pi^+ + \nu \bar{\nu})\lesssim {\cal O}(10^{-10})$, consistent with the SM prediction BR$(K^+ \to \pi^+ + \nu \bar{\nu})|_{\rm SM} \simeq \L 8.4 \pm 1.0 \R \times 10^{-11}$~\cite{SMBG-1,SMBG-2,SMBG-3}. On the other hand, the well-known Grossman-Nir (GN) bound established by the isospin symmetry \cite{GNbound} yields the upper bound
\begin{equation}\label{GN}
{\rm BR}(K_L \to \pi^0 +\nu\bar \nu) < 4.3~{\rm BR}(K^+ \to \pi^++ \nu\bar \nu)|_{\rm exp}.
\end{equation}
As a result, the solution via direct enhancement is ruled out. 

Taking into the different experimental setups, the GN bound can be evaded if one alternatively interprets missing energy as an invisible light particle~\cite{BGN,BGN1,KOTO:2body,Dev:2019hho,Ziegler:2020ize,He:2020jly}. The NA62 collaboration is searching for in-flight decay $K^+ \to \pi^+ \nu \bar{\nu}$, with $\pi^+$ identification and $\gamma$-rejection. The kinematic selection at NA62 leaves a loophole for $m_{Z'}$ close to $m_{\pi^0}$: When the  invariant mass of the invisible particle $m_{miss}$ falls in the interval $[100, 165]$ MeV, the signals suffer from the large background $K^+ \to \pi^+ \pi^0$ which has branching ratio about 21\% of $K^+$ decay~\footnote{The interval $[260, 453]$ MeV is not taken into account neither because of the sizable backgrounds $K^+ \to \pi^+\pi^0 \pi^0$, but it is beyond the interested $m_{Z'}$ mass region for $(g-2)_\mu$. }, so the analysis drops the data. While E949 searches for $K^+ \to \pi^+ \nu \bar{\nu}$ with $K^+$ at rest, and kinematically excludes the interval $[116, 152]$ MeV. 

If $m_{Z'}$ is very close to the neutral pion mass, says $m_{Z'}=m_{\pi^0}\pm \Delta m_{\pi^0}$ with $\Delta m_{\pi^0}\approx 3.8~{\rm MeV}$ the experimental resolution of $\pi^0$ mass, $Z'$ will be constrained by another NA62 analysis~\cite{NA}. This one aims at the invisible decay of $\pi^0$~\cite{NA}, requiring $m_{miss}^2=m_{\pi^0}^2$. It gives the 90\% C.L. upper bound BR$(\pi^0\to {\rm invisible})<4.4\times 10^{-9}$~\cite{CortinaGil:2020zwa} in turn 
\begin{align}\label{pi0:invisible}
{\rm BR}(K^+\to\pi^+Z'\L\to {\rm invisible}\R )<0.9\times 10^{-9} \quad (m_{Z'}=m_{\pi^0}).
\end{align}
A similar strong bound is available from E949. Hereafter, we refer to the loophole region of $m_{Z'}$ with the neighborhood of $m_{\pi^0}$ indicated above removed. In our model, merely $Z'$ in this region is allowed to account for the three KOTO events; maybe only two events can be explained in terms of the analysis in Ref.~\cite{Liao:2020boe}.

\section{The spin-1 candidate for the KOTO anomaly}
\label{sec:rev}

If the KOTO events are robust, then it is a clear signature of light dark world. 
So, it is worth building models which furnish a natural explanation to the KOTO events. 
In this way, we will introduce our model.

\subsection{The profile of $Z'$-induced FCNC for KOTO}

The $s \to d$ transition hinted by KOTO can happen either at tree level via FCNCs in the down-type quark sector or at loop level due to FCNCs originating from the up-type quark sector. 
Alternatively, new physics does not introduce extra FCNCs, and that transition is proceeding in the framework of CKM theory. 
The simplest candidate, a spin-0 scalar mixing with the SM Higgs doublet is such one. 
Nevertheless, its spin-1 similarity, the dark photon does not work~\cite{Jho:2020jsa}. 
In other words, for a light massive gauge boson $Z'$, additional FCNCs beyond the SM is indispensable. 

To that end, as a simple consideration, we presume that $Z'$ comes from a gauged Abelian flavorful symmetry $U(1)_X$ which has the following features:
\begin{itemize}
\item
The SM fermions carry non-universal charges of $U(1)_X$, which then may result in non-simultaneous diagonalization of quark mass matrix and quark-$Z'$ current couplings. Obviously, quarks should be charged under this gauge group. 
\item
Besides, in order to make $Z'$ dominantly decay into a pair of invisible particles, neutrinos or dark matter-like states are also supposed to be charged under it. Considering the lightness of $Z'$, we do not need a hierarchy of charges as long as the coupling to electron is suppressed. 
\item
The gauge coupling is tiny, in particular for the case that the tree-level FCNCs are in the down-type quark sector. However, the massive gauge boson is at the sub-GeV level, and hence the spontaneously breaking scale of $U(1)_X$ is high. Therefore, in general there is no light flavon associated with $U(1)_X$. 
\end{itemize}
Model building can be explored along a variety of lines, and in this paper we take advantage of a model proposed by us before~\cite{Kang:2019vng}, which naturally fits the outlined profiles.

\subsection{The local $(B-L)_{23}$ model and its patterns of FCNCs}

Originally, this model aims at addressing the long-standing $(g-2)_\mu$ anomaly via the light $Z'$ from the flavored local $B-L$ extension to the SM; under this gauge group, only the second and third generations of fermions are charged. 
This gauge group is dubbed as $(B-L)_{23}$ in this paper. 
Such an arrangement leads to an electron and proton phobic $Z'$, which helps avoid the relevant strong exclusions such as Borexino~\cite{Borexino,Borexino1,Borexino2} and COHERENT~\cite{Akimov:2015nza,Akimov:2017ade,Akimov:2020pdx,Akimov:2020czh}, thus allowing the desired $Z'$ having a mass $\sim {\cal O}(10)$~MeV and a moderately small gauge coupling $g_{B-L}\sim {\cal O}(10^{-4} \mathchar`- 10^{-3})$. 
Because the $Z'$ has mass below $2m_\mu$ and moreover has suppressed coupling to electron through the kinetic mixing between $Z'$ and the photon, the dominant decay channel is into a pair of neutrinos, having decay width
\begin{equation}\label{life}
\Gamma(Z')\approx \frac{g_{B-L}^2}{24 \pi} m_{Z'}.
\end{equation}
Then, the lifetime becomes $c\tau_{Z'}\simeq 1.5 \times 10^{-5} \times \L\f{10^{-4}}{g_{B_L}}\R^2\L\f{0.1 \rm GeV}{m_{Z'}}\R $ m. 
The KOTO detector size is $L=3$m, while the size of the NA64 detector is much larger, $L=150$m. 
Since $Z'$ here is invisible, its lifetime is irrelevant to our following discussions. 

To generate the correct CKM structure, additional FCNCs associated with $Z'$ are unavoidable in this model. Therefore, qualitatively this $Z'$ fits the invisible light particle explanation to the KOTO events. If it succeeds quantitatively, accounting for both $(g-2)_\mu$ and the KOTO anomalies at the same time, then the model should deserve top priority. 
Unfortunately, without new CP sources, we will find that the resulting $s \to d$ transition rate is too large for the typical $Z'$ parameters for $(g-2)_\mu$. 
However, our $Z'$ is still a good spin-1 candidate for KOTO, as long as we abandon its responsibility in $(g-2)_\mu$~\footnote{However, hopefully, $(g-2)_\mu$ can be explained by vector-like leptons which are introduced to produce the correct Pontecorvo-Maki-Nakagawa-Sakata matrix in the neutrino sector, even when there are no new CP sources. We leave this to a further publication.}. 

Let us discuss more on the FCNCs in this model. 
The $(B-L)_{23}$ forbids the mixings between the first and other two generations of fermions. 
Introducing flavons to regenerate these mixings then leads to FCNCs. 
Its patterns depend on the origins of the mixings, from the up- and/or down-type quark sectors. 
As a matter of fact, for the case that there are FCNCs in the down-type quark sector, in Ref.~\cite{Kang:2019vng} we have already taken into the constraints from the KOTO report which has not claimed the excess yet~\cite{KOTO}. 
So, one can simply utilize the results there to derive the viable parameter space for three events. 

We focus on the case that the FCNCs are present only in the up-type quark sector, described by the following terms
\begin{eqnarray}\label{FCNCs}
-{\cal L}^u_{Z'}&=&  \bar{u}_{i} \gamma^{\mu}\left[\left(g_{L}^{u}\right)_{i j} P_{L}+\left(g_{R}^{u}\right)_{i j} P_{R}\right] u_{j} Z_{\mu}^{\prime},
\end{eqnarray}
where the Hermitian coupling matrices are defined as
\begin{align}\label{gLR}
(g^u_L)_{ij} &= \frac{g_{B-L}}{3} \Bigl[ \delta_{ij} - (U_u)_{1i}^{\ast} (U_u)_{1j} \Bigr],
\\
(g^u_R)_{ij} &= \frac{g_{B-L}}{3} \Bigl[ \delta_{ij} - (W_u)_{1i}^{\ast} (W_u)_{1j} \Bigr].
\end{align}
where $g_{B-L}$ is the gauge coupling; $U_u$ and $W_u$ are diagonalizing matrices of up quark Yukawa coupling $Y_u$ for left- and right-handed fields, respectively: $Y_u^{\rm diag} = U_u^{\dagger} Y_u W_u$. Note that in the above expressions the parts giving rise to FCNCs are determined by $(U_u)_{1i}$ and  $(W_u)_{1i}$, which is traced back to the fact the FCNCs originate from the first and other two families of fermions carrying different $B-L$ charge. 

Using the above feature, and working in the favored scenario which takes advantage of a singlet flavon plus up-quark-like vector-like fermions to realize CKM, one can show that the coupling matrix $g^u_L$ can be completely determined by the CKM elements, up to $g_{B-L}$. The CKM matrix is defined by the mixing matrices for the left-handed quarks, 
\begin{align}
 \quad V_{\rm CKM} &= U_u^{\dagger} U_d = \begin{pmatrix}
V_{ud} & V_{us} & V_{ub} \\
V_{cd} & V_{cs} & V_{cb} \\
V_{td} & V_{ts} & V_{tb}
\end{pmatrix} \nonumber \\[0.5ex]
&\simeq
\begin{pmatrix}
0.974 & 0.225 & 0.00118 - 0.00345 i \\
- 0.224 - 0.000142 i & 0.974 - 0.0000326 i & 0.0421 \\
0.00831 - 0.00335 & - 0.0413 - 0.000773 & 0.999
\end{pmatrix}.
\label{eq:CKM}
\end{align}
Moreover, $U_d$ is block diagonalized and can be parameterized as
\begin{align}
U_d = \begin{pmatrix}
1 & 0 & 0 \\
0    \\
0   
\end{pmatrix},
\label{eq:ULd}
\end{align}
where the blank block denotes the $2\times 2$ unitary matrix used to diagonalize the second and third families of down-type quarks; it may contain some CP phases, but dull in the FCNC processes in studying. By substituting Eq.~\eqref{eq:ULd} for Eq.~\eqref{eq:CKM} one obtains
\begin{align}\label{eq:CKM1}
(U_u)_{1i} = (U_d V_{\rm CKM}^{\dagger})_{1i} = (V_{ud}^{\ast}, V_{cd}^{\ast}, V_{td}^{\ast}).
\end{align}
As a result, we obtain the explicit numerical form of $g^u_L$ in terms of the CKM elements:
\begin{align}\label{gLu}
g^u_L\simeq g_{B-L} \begin{pmatrix}
0.017 & 0.073 - 0.000046 i & - 0.0027 - 0.0011 i \\
0.073 + 0.000046 i & 0.32 & 0.00062 + 0.00025 i \\
- 0.0027 + 0.0011 i & 0.00062 - 0.00025 i & 0.33
\end{pmatrix}.
\end{align}
The diagonal elements of $g_L^u$ are real (true also for $g_R^u$), and the suppression of  $(g^u_L)_{11}$ is a result of the neutrality of the first generation fermions under the $B-L$ group. In particular, the largest CP violation is from the $(1,3)$- and $(3,1)$-element, $\sim 10^{-3}$, with others suppressed by orders of magnitude. 

On the contrary, the structure of $W_u$ cannot be determined as $U_u$ in Eq.~\eqref{eq:CKM1} since there is no relation like in Eq.~\eqref{eq:CKM}. In principle it is regarded as a generic three by three unitary matrix, and in the later discussions we will make a detailed study on $(W_u)_{1i}$, to investigate its impacts on the meson rare decays. Of interest, $W_u$ can introduce some new CP sources which will largely contribute to $K_L \to \pi^0 Z'$ decay and has the potential to reduce its width by cancellation. 

The readers may wonder if there are other advantages of the gauge group chosen here, since merely arranging the second or the third generation of fermions charged under $B-L$ basically leads to a $Z'$ assembling this one. 
A strong support may be from neutrino physics. Letting the second and third generations of fermions charged under $B-L$ gives a better understanding on neutrino masses and mixings: Two right handed neutrinos are necessary to cancel anomalies, which is the minimal number to produce the acceptable neutrino mass pattern in the seesaw mechanism; moreover, the gauge symmetry leads to the approximate $\mu-\tau$ symmetry demonstrated in neutrino mixings.

\section{$K \to \pi Z'$ from up-type quark FCNC insertion}
\label{sec:calc}

In our last study, we merely studied the FCNCs in the up-type quark sector given in Eq.~(\ref{FCNCs}), e.g., the top quark rare decay $t\to cZ'$, but we neglected the induced FCNCs in the down-type quark sector via the $W$-loop. They are the targets of this paper, and we will first calculate $K \to \pi Z'$ and then investigate its implications to the model, facing the KOTO anomaly and as well the null results from E949/NA62.

\subsection{Calculation of $K \to \pi Z'$}

\begin{figure}[t]
\begin{center}
\includegraphics[width=0.6\textwidth,bb= 0 0 626 345]{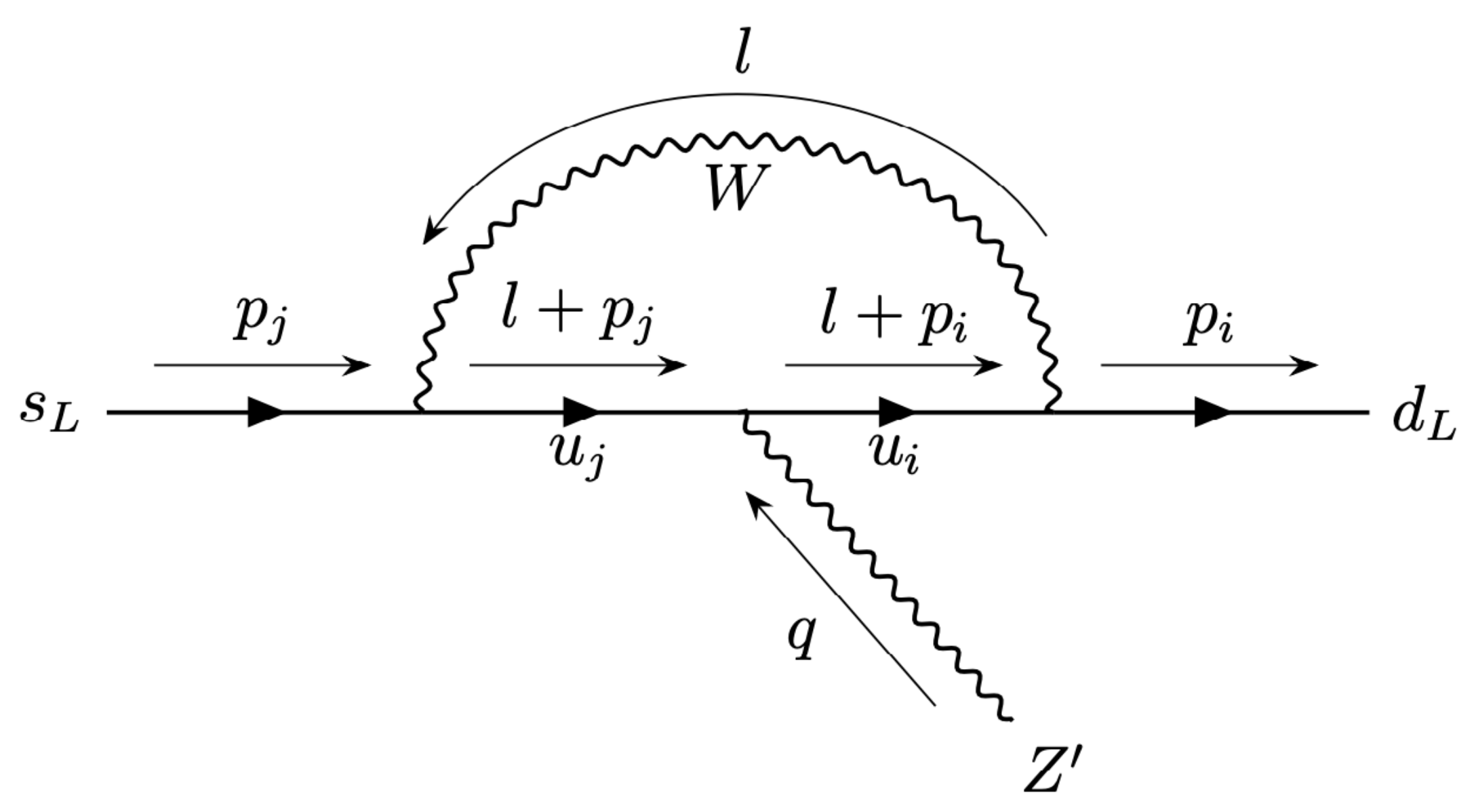}
\end{center}
\caption{1-loop diagram from FCNC coupling of $Z'$ in up-type quark sector. Each momentum is defined here.}
\label{stod}
\end{figure}
At quark level, this process is described by the effective vertex $\bar{d}(p_i)  \Gamma^{\mu}(p_i,p_j)  s(p_j) $, and by taking advantage of the Lorentz invariance and Ward-Takahashi identity one reaches the following structure (up to possible chiral projection operators)
\begin{align}\label{EFT}
\Gamma^{\mu}(p_i,p_j)\sim A\gamma^\mu +B \L q^2\gamma^\mu-\slg{q}q^\mu\R+C\sigma^{\mu\nu}q_\nu,
\end{align}
where the coefficients are functions of $q^2$ with $q=p_j-p_i$ the momentum carried by $Z'$. We will not give a complete calculation of $\Gamma^{\mu}(p_i,p_j)$ which involves a couple of Feynman diagrams. Instead,  here we just concentrate on the dominant one which is shown in Fig.~\ref{stod}, the $Z'$-penguin diagram.  Its contribution then is read from 
\begin{align}
\frac{g_2^2}{2} V_{u_i d}^{\ast} V_{u_j s} \bar{d}(p_i) &\int \! \! \frac{d^4 l}{(2 \pi)^4} \gamma^{\nu} P_L \frac{\slg{l} + \slg{p}_i + m_i}{(l + p_i)^2 - m_i^2} \gamma^{\mu} \left[ (g_L^u)_{ij} P_L + (g_R^u)_{ij} P_R \right] \nonumber \\
&\hspace{9.0em} \times \frac{\slg{l} + \slg{p}_j + m_j}{(l + p_j)^2 - m_j^2} \gamma^{\rho} P_L s(p_j) \frac{g_{\nu \rho}}{l^2 - m_W^2} Z'_{\mu},\nonumber
\end{align}
where $g_2$ is the $SU(2)_L$ gauge coupling, and $m_i$ is the mass of $i$-th generation of up-type quark; $V_{ij}$ is the $(i,j)$ element of the CKM matrix, containing the SM flavor violations in the charged current. We further approximate the masses of the down and strange quarks to be zero. It leads to the vanishing dipole terms in Eq.~(\ref{EFT}), $C\ra 0$, because such terms require chirality flip, namely $C\propto m_{s/d}/m_W^2$. Moreover, the $\slg{q}q^\mu$ term automatically vanishes after using the motion of equations for the fermions. Therefore, we expect that the $Z'$-penguin diagram leads to an effective coupling $g_{dsZ'}^{\rm eff}(q^2)\bar{d}(p_i) \gamma^{\mu} P_L s(p_j)$. 

Now let us calculate $g_{dsZ'}^{\rm eff}(q^2)$ explicitly, using the public codes, FeynCalc \cite{FeynCalc1,FeynCalc2,FeynCalc3}~\footnote{\href{https://feyncalc.github.io/}{https://feyncalc.github.io/}} and LoopTools \cite{LoopTools}~\footnote{\href{http://www.feynarts.de/looptools/}{http://www.feynarts.de/looptools/}}. The result of loop function from FeynCalc is
\begin{align}
\bar{d}(p_i) &\int \! \! \frac{d^4 l}{(2 \pi)^4} \gamma^{\nu} P_L \frac{\slg{l} + \slg{p}_i + m_i}{(l + p_i)^2 - m_i^2} \gamma^{\mu} \left[ (g_L^u)_{ij} P_L + (g_R^u)_{ij} P_R \right] \frac{\slg{l} + \slg{p}_j + m_j}{(l + p_j)^2 - m_j^2} \gamma^{\rho} P_L s(p_j) \frac{g_{\nu \rho}}{l^2 - m_W^2} \nonumber \\
&= - \frac{2}{16 \pi^2} \Bigl\{ (g^u_L)_{ij} \left[ q^2 ( C_0 + C_1 + C_2 + C_{12} ) - 2 C_{00} \right] + (g^u_R)_{ij} m_{u_i} m_{u_j} C_0 \Bigr\} \bar{d}(p_i) \gamma^{\mu} P_L s(p_j),
\label{eq:loopint}
\end{align}
where $C_a$ $(a = 0, 1, 2, 00, 12)$ are Passarino-Veltman (PV) integrals \cite{PV}. 
Since we have taken $m_{d, s} \to 0$, the arguments for the PV integrals are reduced to
\begin{align}
C_0 : C_0 (0, 0, q^2, m_i^2, m_W^2, m_j^2) \,\, \text{and} \,\, C_a : C_a (0, q^2, 0, m_W^2, m_i^2, m_j^2) \,\, (\text{for } a = 1, 2, 00, 12).
\end{align}
Among them, only $C_{00}$ does not scale as $1/m_W^2$ thus dominant in the effective coupling. One can gain more insights into the scaling behavior of the PV integrals by developing approximations like in Ref.~\cite{Baek:2015fma}.  The effective coupling is given by
\begin{align}\label{geff}
g_{dsZ'}^{\rm eff} &\equiv  - \frac{g_2^2}{16 \pi^2} \sum_{i,j =1}^3 V_{u_i d}^{\ast} V_{u_j s} \Bigl\{ (g^u_L)_{ij} \left[ q^2 ( C_0 + C_1 + C_2 + C_{12} ) - 2 C_{00} \right] + (g^u_R)_{ij} m_{u_i} m_{u_j} C_0 \Bigr\},
\end{align}
which depends not only on $(g^u_L)_{ij}$ but also on $(g^u_R)_{ij}$, and note that generically both of them are complex. It is convenient to rewrite 
\begin{align}
g_{dsZ'}^{\rm eff} = - \frac{1}{16 \pi^2} \sum_{i,j =1}^3 \Bigl[ (C_L^{ds})_{ij} (g^u_L)_{ij} + (C_R^{ds})_{ij} (g^u_R)_{ij} \Bigr],
\label{eq:gdseff}
\end{align}
where $C_{L,R}^{ds}$ are the combinations of CKM elements and PV integrals specified in Eq.~(\ref{geff}). 

With the effective vertex, now we can calculate the branching ratios for $K \to \pi Z'$ processes by using the following results~\cite{Fuyuto:2014cya,Fuyuto:2015gmk}:
\begin{align}
{\rm BR} (K^+ \to \pi^+ Z') &= \frac{| g_{dsZ'}^{\rm eff} |^2}{64 \pi} \frac{\lambda \left(m_{K^+}^2, m_{\pi^+}^2, m_{Z'}^2 \right)^{3/2}}{m_{Z'}^2 m_{K^+}^3 \Gamma_{K^+}} \Bigl[ f_+^{K^+ \pi^+} \left( m_{Z'}^2 \right) \Bigr]^2, \label{eq:BRKp} \\
{\rm BR} (K_L \to \pi^0 Z') &= \frac{\left( {\rm Im} \, g_{dsZ'}^{\rm eff} \right)^2}{64 \pi} \frac{\lambda \left(m_{K_L}^2, m_{\pi^0}^2, m_{Z'}^2 \right)^{3/2}}{m_{Z'}^2 m_{K_L}^3 \Gamma_{K_L}} \Bigl[ f_+^{K^0 \pi^0} \left( m_{Z'}^2 \right) \Bigr]^2, \label{Brs}
\end{align}
where $m_K$ and $\Gamma_K$ are mass and decay width of kaon, respectively; $\lambda (x, y, z) = x^2 + y^2 + z^2 - 2 x y - 2 y z - 2 z x$, and $f_+^{K \pi} (q^2)$ is the $K \to \pi$ form factor \cite{Mescia:2007kn}. 

Remarkably, the charged kaon decay proceeds without CP violation and BR$\left(K^{+} \rightarrow \pi^{+} Z^{\prime}\right)\propto \left|g_{d s Z^{\prime}}^{\mathrm{eff}}\right|^{2}$, whereas the neutral kaon decay requires it and BR$\left(K_{L} \rightarrow \pi^{0} Z^{\prime}\right)$ is proportional to the squared imaginary part of the effective coupling. In the SM, CP violation is known to be small, and therefore in general  BR$\left(K_{L} \rightarrow \pi^{0} Z^{\prime}\right)$ is supposed to be at least moderately suppressed.

\subsection{Analysis on $C^{ds}_{L,R}$}

To develop the numerical impression on $(C_{L,R}^{ds})$, we set $q^2 (= m_{Z'}^2) = m_{\pi^0}^2$ as a reference value, and then one obtains
\begin{align}
C_L^{ds} &= \begin{pmatrix}
- 0.34 + 2.7 \times 10^{-7} i & - 1.5 + 4.9 \times 10^{-5} i & 7.9 \times 10^{-2} + 1.5 \times 10^{-3} i \\
7.8 \times 10^{-2} - 4.9 \times 10^{-5} i & 0.34 - 2.2 \times 10^{-4} i & - 1.8 \times 10^{-2} - 3.3 \times 10^{-4} i \\
- 3.7 \times 10^{-3} - 1.5 \times 10^{-3} i & - 1.6 \times 10^{-2} - 6.4 \times 10^{-3} i & 7.2 \times 10^{-4} + 3.1 \times 10^{-4} i
\end{pmatrix}, \label{eq:CL} \\
C_R^{ds} &= \begin{pmatrix}
9.6 \times 10^{-10} + 2.2 \times 10^{-10} i & 1.5 \times 10^{-6} - 4.9 \times 10^{-11} i & - 4.3 \times 10^{-7} - 8.0 \times 10^{-9} i \\
- 7.7 \times 10^{-8} + 4.9 \times 10^{-11} i & - 1.7 \times 10^{-4} + 1.1 \times 10^{-7} i & 5.7 \times 10^{-5} + 1.0 \times 10^{-6} i \\
2.0 \times 10^{-8} + 8.0 \times 10^{-9} i & 4.9 \times 10^{-5} + 2.0 \times 10^{-5} i & - 1.1 \times 10^{-4} - 4.5 \times 10^{-5} i
\end{pmatrix},
\label{eq:CR}
\end{align}
where the masses of up-type quark and the CKM matrix are taken from PDG 2019 \cite{PDG2019}. Several observations are in orders: 
\begin{itemize}
\item
It is clear that for most elements the size of $(C_L^{ds})_{ij}$ is much larger than that of $(C_R^{ds})_{ij}$, due to the fact that the former receives the $C_{00}$ contribution.  $(C_R^{ds})_{33}$ is an exception, because it benefits from the $m_t^2$ enhancement. 
\item
$| (C_L^{ds})_{12} |$, without involving flavor violation from the charged current, is the largest element as expected and would be the dominant contribution to $s \to d Z'$ processes. 
\item  It is notable that some elements of $C_R^{ds}$, in particular $(C_R^{ds})_{32,23}$ and $(C_R^{ds})_{33}$, have comparable size with those of $C_L^{ds}$. Hence, these may contribute to $K_L \to \pi^0 Z'$ decay process, depending on the size of $(g_R^u)_{ij}$, namely, the structure of $W_u$. 
\end{itemize}
The last feature motivates us to consider two scenarios: I) omit contributions of $(C_R^{ds})_{ij} (g_R^u)_{ij}$ with $i, j$ summed~\footnote{Actually, the contributions from $(C_R^{ds})_{ij} (g_R^u)_{ij}$ cannot be omitted in any structure of $W_u$ since some elements in $(g_R^u)_{ij}$ still exist in our model. However, as long as we discuss the prediction of $K_L \to \pi^0 Z'$, we can ignore its contributions by setting appropriate structure of $W_u$.}; II) include contributions of $(C_R^{ds})_{ij} (g_R^u)_{ij}$. It is interesting that for the Scenarios I, our predictions on $K\ra\pi Z'$ can be explicitly determined by the SM parameters except for $g_{B-L}$. In this sense, the  Scenario I corresponds to the model in which there are no new CP violation sources. On the other hand, the CP violation in the Scenario II is not completely determined by the SM parameters, owing to the arbitrariness of $g_R^u$. This additional CP violation may admit an elaborate cancellation between Im$[(C_R^{ds})_{ij} (g_R^u)_{ij}]$ and Im$[(C_L^{ds})_{ij} (g_L^u)_{ij}]$, thus opening the possibility to explain both of $(g-2)_{\mu}$ and KOTO anomalies in our model.

\subsection{Implications to the model}

In this subsection we investigate the implications of induced $K\ra \pi+Z'$ to the local $(B-L)_{23}$ model in two scenarios, with Scenario I simply dropping the contribution from $(C_R^{ds})_{ij} (g_R^u)_{ij}$ for illustration, while Scenario II highlighting its additional CP violation. We will find that, in general the KOTO events can be easily explained in the Scenario I if we give up the original motivation, to account for the $(g-2)_\mu$ discrepancy. Otherwise, we should fall back on the other scenario.

\subsubsection{Scenario I: Close the door for $(g-2)_\mu$ but open the window for KOTO}

We first discuss the scenario where the contributions from $(C_R^{ds})_{ij} (g_R^u)_{ij}$ is ignored. In this limit, the strong exclusion on the $(m_{Z'},g_{B-L})$ parameter plane from $K$ rare decays is clear, so it is questionable that if the remaining parameter space that is capable of accounting for $(g-2)_\mu$ survives. 

For illustration, let us choose a point characterized by $Z'$ mass very close to $m_{\pi^0}$, e.g., $m_{Z'} = m_{\pi^0}$ and $g_{B-L} = 10^{-3}$ to explain the $(g-2)_{\mu}$ anomaly. The resulting branching ratio for $K_L \to \pi^0 Z'$ is
\begin{align}
{\rm BR} (K_L \to \pi^0 Z')_{\rm our} = 3.4 \times 10^{-6},
\end{align}
which is much larger than the measured value, BR$(K_L \to \pi^0 Z') = \mathcal{O}(10^{-9})$. Therefore, this example point must have been excluded by KOTO. For this $m_{Z'}$, the constraint in Eq.~(\ref{pi0:invisible}) applies and imposes an even stronger bound
\begin{align}
g_{B-L} < 5.4 \times 10^{-6}.
\label{eq:Kpconst}
\end{align}
The bound from E949 experiment at 90\% C.L.~\cite{E949-2} is much weaker, $g_{B-L} < 4.2 \times 10^{-5}$. Nevertheless, in the loophole region, e.g., $m_{Z'} = 128$ MeV, the $Z'$ can readily explain the KOTO anomaly for
\begin{align}
1.6 \times 10^{-5} \ (1.0 \times 10^{-5}) \lesssim g_{B-L} \lesssim 3.3 \times 10^{-5} \ (4.0 \times 10^{-5}),
\end{align}
within 1$\sigma$ (2$\sigma$) error. 
 
\begin{figure}[t]
\begin{center}
\includegraphics[width=0.6\textwidth,bb= 0 20 450 452]{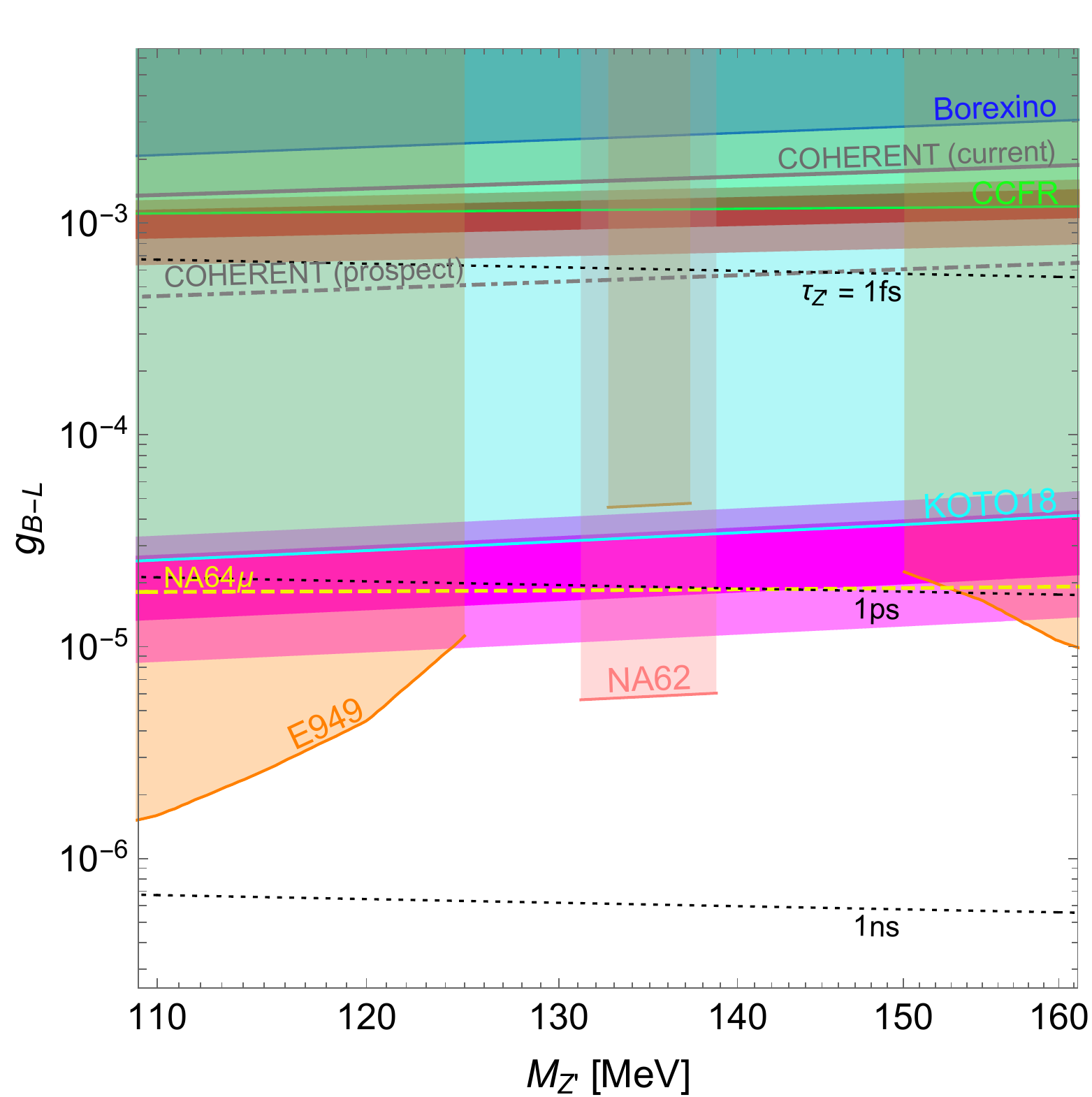}
\end{center}
\caption{Parameter space for $(g-2)_{\mu}$ and KOTO result in the Scenario I. The red and magenta band show the favored region for $(g-2)_{\mu}$ and KOTO result at 1$\sigma$ (darker) and at 2$\sigma$ (lighter), respectively. The other shaded regions are excluded by experiments: Borexino (blue)~\cite{Borexino,Borexino1,Borexino2}, COHERENT (gray)~\cite{Akimov:2015nza,Akimov:2017ade,Akimov:2020pdx,Akimov:2020czh}, CCFR (green)~\cite{Mishra:1991bv}, E949 (orange)~\cite{E949-1,E949-2}, KOTO before the events (cyan) \cite{KOTO} and NA62 (pink)~\cite{NA}. The dashed yellow and dashed-dotted gray lines are the future prospects of NA64$\mu$ with $10^{12}$ muons~\cite{Gninenko:2014pea} and COHERENT, respectively. The dotted lines show the contours for the life time of $Z'$, calculated from Eq.~(\ref{life}).}
\label{fig:para}
\end{figure}

The summary plot for the parameter spaces for $(g-2)_{\mu}$ (red band) and KOTO result (magenta band) in the Scenario I is shown in Fig.~\ref{fig:para}. 
%%%%%%%%%%%%%%%%%%%%%%%%%%%%%%%%%%%%%%%%
%%%%%%%%%%%%%%%%%%%%%%%%%%%%%%%%%%%%%%%%
The darker and lighter bands show the favored region at 1$\sigma$ and at 2$\sigma$, respectively. Other shaded regions are excluded by these experiments: Borexino (blue) \cite{Borexino,Borexino1,Borexino2}, COHERENT (gray)~\cite{Akimov:2015nza,Akimov:2017ade,Akimov:2020pdx,Akimov:2020czh}~\footnote{This constraint is obtained mainly from the muon neutrino source with interactions with up and down quarks in nucleon. In Ref.~\cite{Altmannshofer:2018xyo}, the authors discuss about the constraints also from the strange quark content in nucleon, which yields a relatively weak bound. 
}, CCFR (green) \cite{Mishra:1991bv}, E949 (orange) \cite{E949-1,E949-2}, KOTO before the events (cyan) \cite{KOTO} and NA62 (pink) \cite{NA}. The dotted lines show the contours for the life time of $Z'$, calculated from Eq.~(\ref{life}). It is seen that, for any value of $m_{Z'}$ inside the loophole region, the required size of $g_{B-L}$ to account for the $(g-2)_{\mu}$ discrepancy is about two orders of magnitude larger than the upper bound by KOTO; outside the loophole, E949 yields the strongest bound and definitely rules out the possibility to explain $(g-2)_{\mu}$~\footnote{Recently, NA62 experiment provides upper bounds on BR($K^+ \to \pi^+ Z'$) for the mass ranges of $m_{Z'} < 110$ MeV and $154 \, {\rm MeV} < m_{Z'} < 260$ MeV~\cite{CortinaGil:2020fcx}. 
We do not show its constraints in Fig.~\ref{fig:para} since it is irrelevant to the following discussion.}. 
In this figure, we also show the future prospect of NA64 with dedicated muon beam, denoted as NA64$\mu$ (dashed yellow). 
This prospect is calculated with $10^{12}$ incident muons~\cite{Gninenko:2014pea}, and its upper bound on $g_{B-L}$ in this mass region is about two orders of magnitude smaller than the required value for $(g-2)_{\mu}$ explanation. 
Interestingly, this prospect can search the parameter space for KOTO results in Scenario I. 
In addition, the future prospect of COHERENT is also shown by dashed-dotted gray line.
Note that the COHERENT constraint and prospect for our model can be translated from the ones for $L_{\mu}-L_{\tau}$ model~\footnote{Its constraint and prospect can be found, for example in Refs.~\cite{Abdullah:2018ykz,Bauer:2018onh,Amaral:2020tga,Cadeddu:2020nbr,Banerjee:2020zvi}.} by considering the difference of the kinetic mixing between two models. 
Although this is weaker than NA64$\mu$ prospect, it's possible to search all parameter space for $(g-2)_{\mu}$ explanation of our model, and we will turn back to this point in the Scenario II.

In the next scenario, we will demonstrate that BR$(K_L \to \pi^0 Z')$ can be significantly reduced and then both the  $(g-2)_{\mu}$ and KOTO anomalies can be explained, at least in the loophole region of $m_{Z'}$.

\subsubsection{Scenario II: One stone for two birds at the price of moderate tuning}

In the Scenario I, the largeness of the branching ratio of $K_L \to \pi^0 Z'$ is due to the large value of Im$[(C_L^{ds})_{ij} (g^u_L)_{ij}]$. For concreteness, from the $g_L^u$ matrix Eq.~(\ref{gLu}) and the $C_L^{ds}$ matrix Eq.~(\ref{eq:CL}), one has Im$[(C_L^{ds})_{ij} (g^u_L)_{ij}]\approx 0.88\times 10^{-5}$; we set $q^2=m_{\pi^0}^2$ for reference unless otherwise specified. However, in the Scenario II by switching on the Im$[(C_R^{ds})_{ij} (g^u_R)_{ij}]$ contribution, there is a possibility to cancel this size by about two orders of magnitude, hence to explain both anomalies. The corresponding fine-tuning of CP violation may be not very serious, since we find that the elements Im$(C_{L/R}^{ds})_{ij} (g^u_{L/R})_{ij}$ (not summed) already accidentally cancel each other out to a degree $\sim 90\%$. In the following we make a detailed discuss on this cancellation. 

As mentioned before, Im$(C_R^{ds})_{32,33}\sim {\cal O}(10^{-5})$ are large enough to contribute to Im$(g_{dsZ'}^{\rm eff})$. Moreover, Re$(C_R^{ds})_{23,32}$ are sufficiently large and they, along with the sizable Im$(g^u_R)_{23,32}$ (namely the CP violation from the corresponding elements of $W_u$), may play an important role in Im$(g_{dsZ'}^{\rm eff})$. In order to understand what is the proper pattern of $W_u$ good for reducing  Im$(g_{dsZ'}^{\rm eff})$, we generate its elements randomly. From Eq.~(\ref{gLR}), the relevant elements are $(W_u)_{1i}$, which in principle are free parameters except for satisfying the unitary condition:
\begin{align}
(W_u)_{1i} = (r_{11} e^{i \theta_{11}}, r_{12} e^{i \theta_{12}}, r_{13} e^{i \theta_{13}}), 
\end{align}
where $| r_{1i} | \leq 1$ satisfying the relation
\begin{align}\label{unitary}
| r_{11} |^2 + | r_{12} |^2 + | r_{13} |^2 = 1.
\end{align} 
 For example, an illustrative choice is
\begin{align}\label{wu1}
(W_u)_{1i} = \left(0, \frac{1}{\sqrt{2}}, \frac{1}{\sqrt{2}} \, e^{i \theta_{13}} \right).
\end{align}
Then, when $m_{Z'} = 128$ MeV and $\theta_{13} \simeq 0.59 \pi$, BR$(K_L \to \pi^0 Z') \simeq 2.1 \times 10^{-9}$ is realized with $g_{B-L} = 10^{-3}$ which is needed to explain the $(g-2)_{\mu}$ anomaly. 

\begin{figure}[t]
\begin{center}
\includegraphics[width=0.33\textwidth,bb= 0 0 420 409]{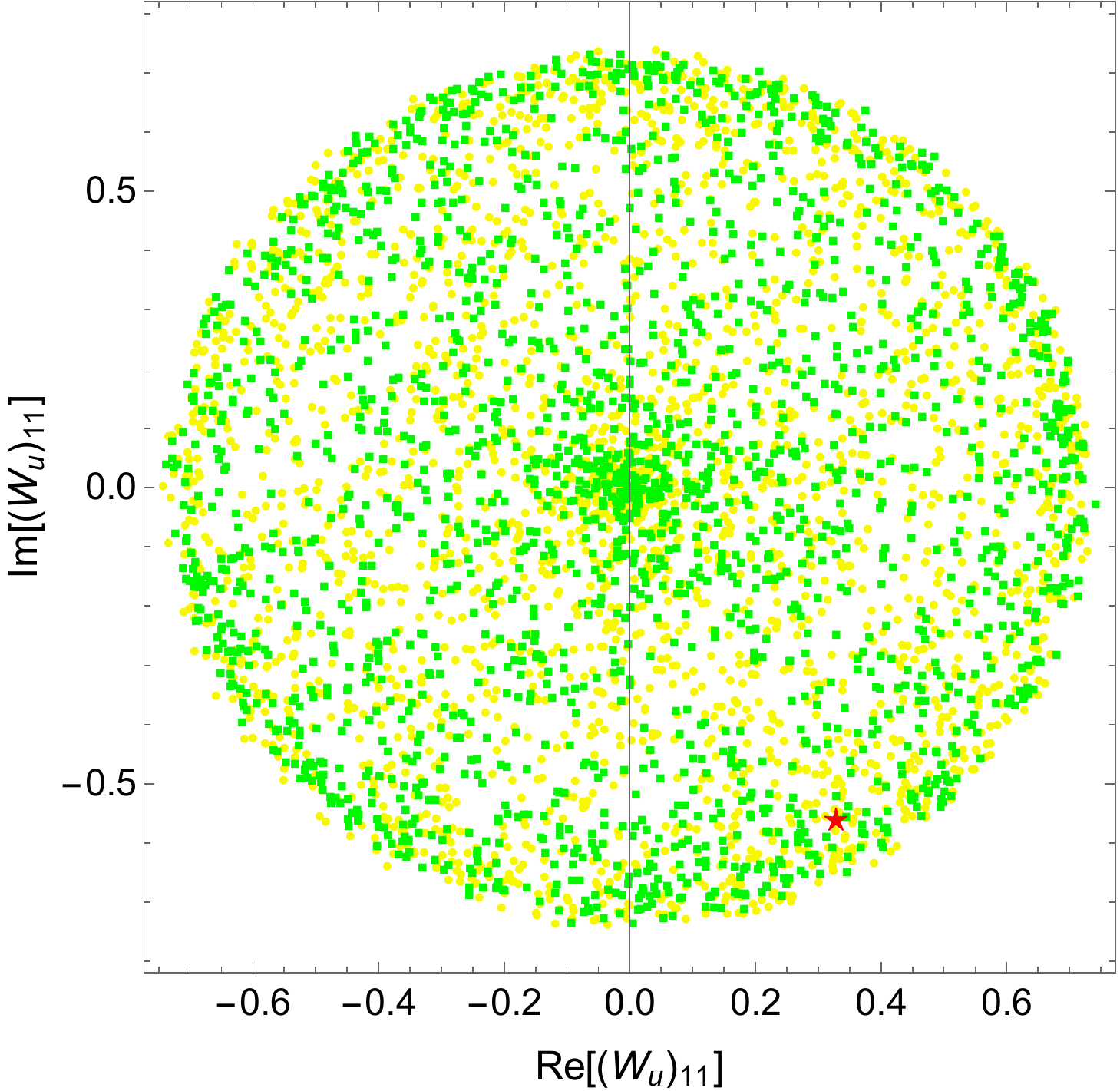}~~\includegraphics[width=0.33\textwidth,bb= 0 0 420 409]{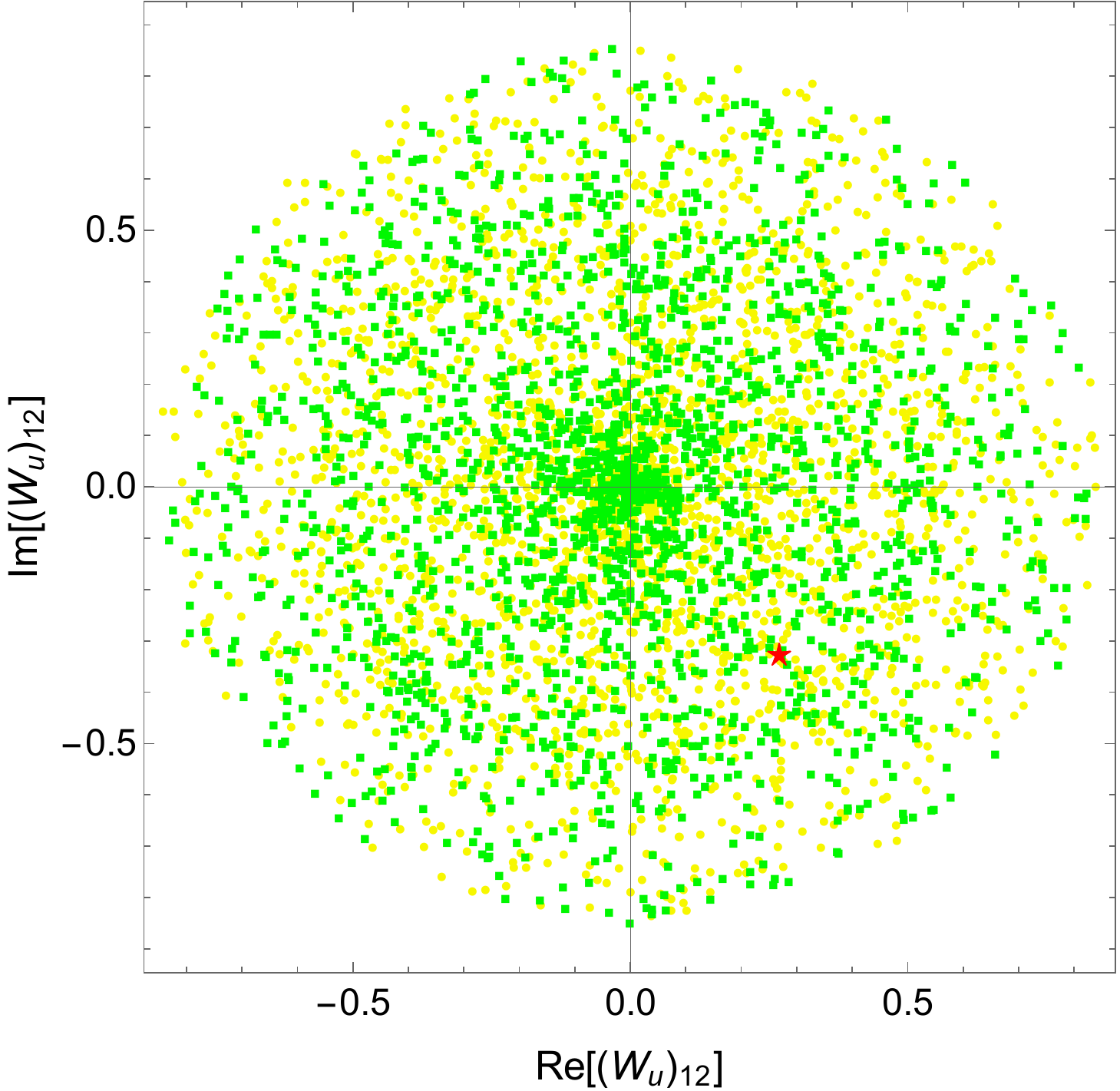}~~\includegraphics[width=0.33\textwidth,bb= 0 0 420 409]{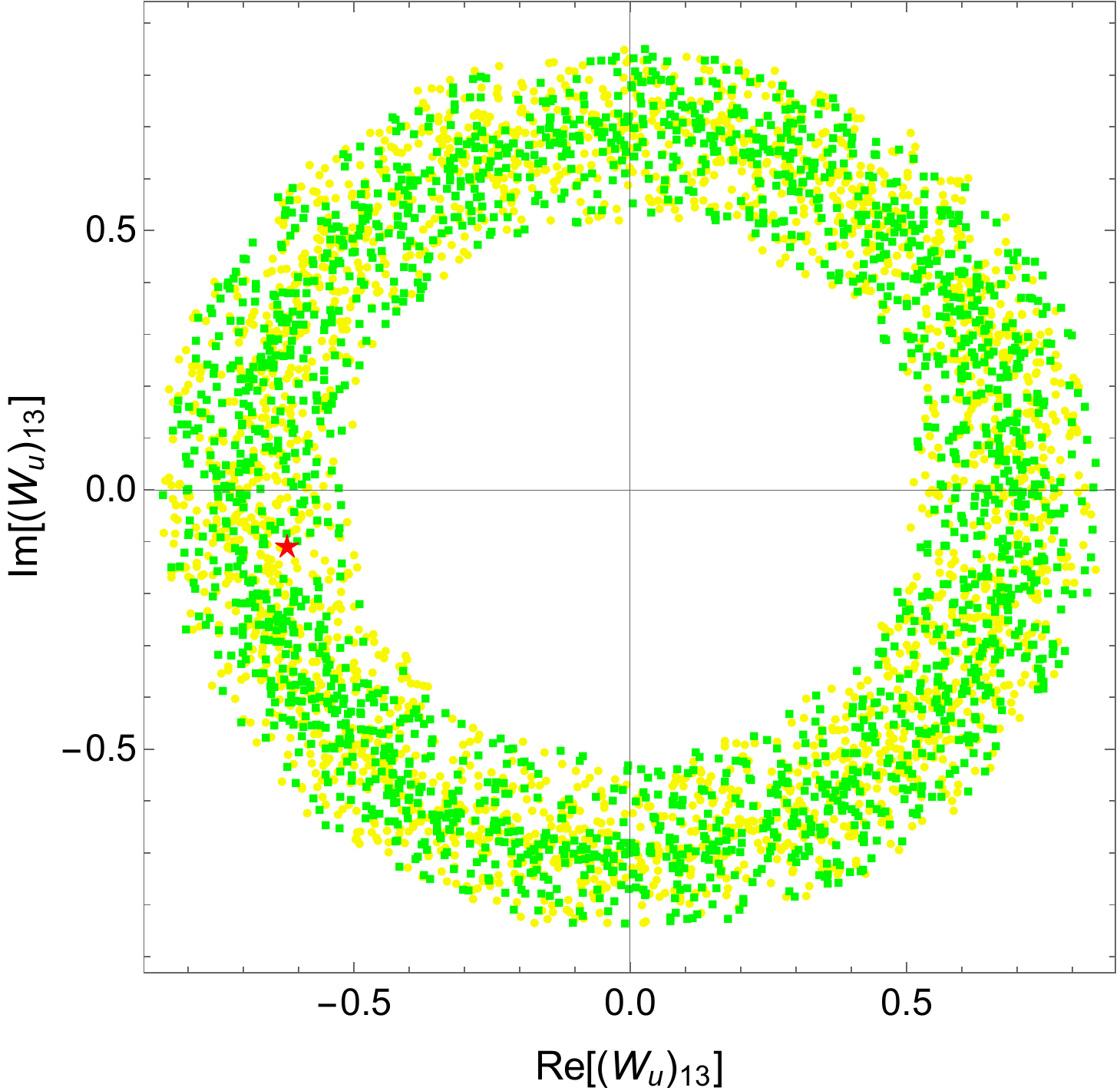}
\end{center}
\caption{Each element in $(W_u)_{1i}$ which can explain $(g-2)_{\mu}$ and KOTO anomalies within $1\sigma$ (green square) and $2\sigma$ (yellow circle). In these plot, we set $m_{Z'} = 140$ MeV. The red star denotes the benchmark point for Fig.~\ref{fig:bench}.}
\label{fig:Wu1i}
\end{figure}

As a general survey, we generate $10^5$ samples for the $(W_u)_{1i}$ elements and check the prediction of the favored $g_{B-L}$ value for $(g-2)_{\mu}$ and KOTO anomalies. We show each element which can explain both anomalies within $2\sigma$ in Fig.~\ref{fig:Wu1i}. 
%%%%%%%%%%%%%%%%%%%%%%%%%%%%%%%%%%%%%%%%
%%%%%%%%%%%%%%%%%%%%%%%%%%%%%%%%%%%%%%%%
The green square and yellow circle denote the points where both anomalies are explained within $1\sigma$ and $2\sigma$, respectively. 
For this figure, we set $m_{Z'} = 140$ MeV, but we find that the similar results are obtained for a different value of $m_{Z'}$ within the loophole regions.

We can observe some important features of these elements. First, all of the elements are bounded from above, $| (W_u)_{1i} | < 0.7 \mathchar`- 0.8$. Then, at least two elements of $(W_u)_{1i}$ are needed to satisfy Eq.~(\ref{unitary}), like the example in Eq.~(\ref{wu1}). Second, $|(W_u)_{11,12}|$ can be small, while $|(W_u)_{13}|$ should be $0.5 \sim 0.8$. The reason is understood by nothing but that Im$(C_R^{ds})_{33}$ tends to be even larger than  Im$[(C_L^{ds})_{ij} (g^u_L)_{ij}]$,  and consequently a sizable $|(W_u)_{13}|$ is necessary to lower down $(g^u_R)_{33}\propto (1-|(W_u)_{13}|^2)$, thus allowing the cancellation to happen. Therefore, we cannot explain both anomalies with $W_u \sim V_{\rm CKM}$, and some different and specific structure for $W_u$ is needed.  Since this specific structure is due to the structure of $C^{ds}_R$ in Eq.~\eqref{eq:CR}, which is obtained only from the SM parameters, the required structure of $W_u$ is specific to our setup. 

\begin{figure}[t]
\begin{center}
\includegraphics[width=0.6\textwidth,bb= 0 0 450 437]{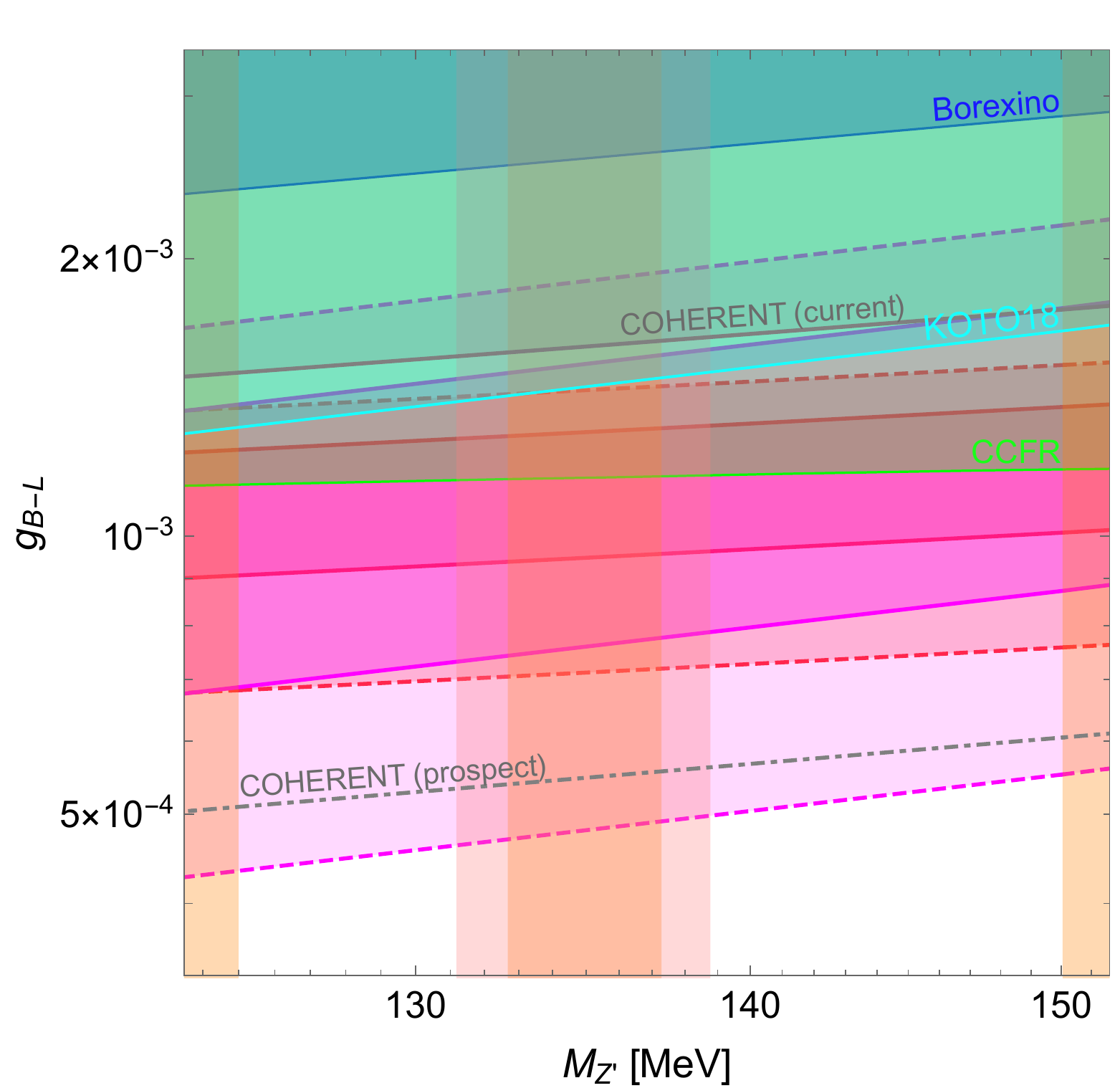}
\end{center}
\caption{Parameter space for $(g-2)_{\mu}$ and KOTO result in the Scenario II, using the benchmark values for $(W_u)_{1i}$ in Fig.~\ref{fig:Wu1i}. The color manner is the same as in Fig.~\ref{fig:para}. In order to specify each band, we change the boundaries for $1\sigma$ and $2\sigma$ to solid and dashed lines, respectively.}
\label{fig:bench}
\end{figure}

In Fig.~\ref{fig:bench}, we show the summary plot for the benchmark point in Fig.~\ref{fig:Wu1i}. 
%%%%%%%%%%%%%%%%%%%%%%%%%%%%%%%%%%%%%%%%
%%%%%%%%%%%%%%%%%%%%%%%%%%%%%%%%%%%%%%%%
The color manner is the same as in Fig.~\ref{fig:para}, but we change the boundaries of each favored band for $1\sigma$ and $2\sigma$ to solid and dashed lines, respectively. 
It is clear that both anomalies can be explained with $g_{B-L} = \mathcal{O}(10^{-3})$, and these parameter space will be searched by the COHERENT experiment. We emphasize that the future prospect of NA64$\mu$ can be also applied to this scenario, whose expected upper bound can be read as $g_{B-L} \lesssim (1.8 \mathchar`- 1.9) \times 10^{-5}$. 
Therefore, we can expect some signal of our model in NA64$\mu$ as well as COHERENT, and moreover, if such signal predicts $g_{B-L} \sim \mathcal{O}(10^{-3}$), the explanation of both anomalies can be done, based on our Scenario II.

\section{Predictions in the $B$ physics}
\label{sec:Bphys}

We have studied the induced FCNCs in the Kaon system, and in particular explored the possibility to explain two anomalies simultaneously in the Scenario II, by means of a large $g_{B-L}$ but a fine-tuned CP violation in the loophole region of $m_{Z'}$. However, the loophole and as well fine-tuning may be not true in the $B$ meson system, and hence it is important to study the accompanied rare decays of the $B$ mesons, e.g., by $B \to K + Z'(\to \nu \bar{\nu})$~\footnote{It is also of interest to study the detect prospect of radiative $B$ decay $B\ra \gamma Z'$ which is recently proposed in Ref.~\cite{Chen:2020szf}}. Then, the $B$-factory may provide a promising way to test it. 
Actually, the Belle data already imposes a constraint. 

In analogy to $s \to d Z'$, the transitions $b \to q Z'$ $(q = d, s)$ are through the effective $q$-$b$-$Z'$ couplings, with the effective couplings given by
\begin{align}
g_{qbZ'}^{\rm eff} &= - \frac{g_2^2}{16 \pi^2} \sum_{i,j =1}^3 V_{u_i q}^{\ast} V_{u_j b} \Bigl\{ (g^u_L)_{ij} \left[ q^2 ( C_0 + C_1 + C_2 + C_{12} ) - 2 C_{00} \right] + (g^u_R)_{ij} m_{u_i} m_{u_j} C_0 \Bigr\} \nonumber \\
&\equiv - \frac{1}{16 \pi^2} \sum_{i,j =1}^3 \Bigl[ (C_L^{qb})_{ij} (g^u_L)_{ij} + (C_R^{qb})_{ij} (g^u_R)_{ij} \Bigr], \label{eq:Bloop}
\end{align}
with $C_{L,R}^{qb}$ again the known matrices at some $q^2$, and the concrete forms at $q^2=m_{\pi^0}^2$ are cast in Appendix.~\ref{app:btoqnumeval}, from which one can see that the most sizable elements are $(C_{L}^{qb})_{13,23}\sim{\cal O}(1)$. These effective couplings are fixed as long as $g_R^u$ or $(W_u)_{1i}$ is chosen to realize the CP violation cancellation in the Scenario II. 

We calculate the $B$ meson decays by the following formulas \cite{Kramer:1991xw,Oh:2009fm}:
\begin{align}
{\rm BR}(B \to P Z') &= \frac{| g_{qbZ'}^{\rm eff} |^2}{64 \pi \kappa_P^2} \frac{\lambda(m_B^2, m_P^2, m_{Z'}^2)^{3/2}}{m_{Z'}^2 m_B^3 \Gamma_B} | f_+^{BP} (m_{Z'}^2) |^2,  \label{BRpseudo} \\[0.6ex]
{\rm BR}(B \to V Z') &= \frac{| g_{qbZ'}^{\rm eff} |^2}{64 \pi \kappa_V^2} \frac{\lambda(m_B^2, m_V^2, m_{Z'}^2)^{1/2}}{m_B^3 \Gamma_B} \left( |H_0^V|^2 + |H_+^V|^2 + |H_-^V|^2 \right),  \label{BRvector} 
\end{align}
where $m_B$ and $\Gamma_B$ are mass and width of $B$ meson, $f_+^{BP} (q^2)$ is $B \to P$ form factor \cite{Ball:2004ye}, and $\kappa_{P,V}^2$ are 1 for $P = \pi^+, K^{0,+}$ and $V = \rho^+, K^{*0,*+}$ or 2 for $P = \pi^0$ and $V = \rho^0$. 
$H_0^V$ and $H_{\pm}^V$ are the helicity amplitudes which are given as
\begin{align}
H_0^V &= - ( m_B + m_V ) A_1^{BV} (m_{Z'}^2) x_{VZ'} + \frac{2 m_V m_{Z'}}{m_B + m_V} A_2^{BV} (m_{Z'}^2) \left( x_{VZ'}^2 - 1 \right), \\
H_{\pm}^V &= ( m_B + m_V ) A_1^{BV} (m_{Z'}^2) \pm \frac{2 m_V m_{Z'}}{m_B + m_V} V^{BV} (m_{Z'}^2) \sqrt{x_{VZ'}^2 - 1},
\label{BtoK*}
\end{align}
where $A_1^{BV} (q^2)$, $A_2^{BV} (q^2)$ and $V^{BV} (q^2)$ are the form factors for $B \to V$ transition \cite{Ball:2004rg}, and $x_{VZ'} \equiv \left( m_B^2 - m_V^2 - m_{Z'}^2 \right) / \left( 2 m_V m_{Z'} \right)$. Note that the above formulas can be used for both neutral and charged $B$ meson decays, and moreover, unlike $K_L\ra\pi^0 Z'$, the former decays do not need CP violation. 

The results with $m_{Z'} = 128$ MeV and $140$ MeV in the Scenario II are summarized in Table~\ref{tab:Bresult}. In the calculation of these branching ratios, we use the benchmark values for $(W_u)_{1i}$ in Fig.~\ref{fig:Wu1i}. 
In addition, as the reference value, $g_{B-L}$ is chosen to realize the central value of the KOTO result, BR$(K_L \to \pi^0 Z') = 2.1 \times 10^{-9}$. 
Note that each $g_{B-L}$ value is satisfied the CCFR constraint. 
%%%%%%%%%%%%%%%%%%%%%%%%%%%%%%%%%%%%%%%%
\begin{table}[thb]
\begin{center}
\begin{tabular}{c|cc}
$(m_{Z'}, g_{B-L})$ & (128 MeV, $1.02 \times 10^{-3}$) & (140 MeV, $1.15 \times 10^{-3}$) \\ \hline \hline
$B^0 \to \pi^0 Z'$ & $8.12 \times 10^{-7}$ & $8.59 \times 10^{-7}$ \\
$B^+ \to \pi^+ Z'$ & $1.75 \times 10^{-6}$ & $1.85 \times 10^{-6}$ \\ \hline
$B^0 \to \rho^0 Z'$ & $1.00 \times 10^{-6}$ & $1.06 \times 10^{-6}$ \\
$B^+ \to \rho^+ Z'$ & $2.16 \times 10^{-6}$ & $2.28 \times 10^{-6}$ \\ \hline
$B^0 \to K^0 Z'$ & $1.44 \times 10^{-2}$ & $1.53 \times 10^{-2}$ \\
$B^+ \to K^+ Z'$ & $1.56 \times 10^{-2}$ & $1.65 \times 10^{-2}$ \\ \hline
$B^0 \to K^{*0} Z'$ & $1.65 \times 10^{-2}$ & $1.75 \times 10^{-2}$ \\
$B^+ \to K^{*+} Z'$ & $1.78 \times 10^{-2}$ & $1.89 \times 10^{-2}$ \\ \hline
\end{tabular}
\end{center}
\caption{Numerical values for branching ratios of $B$ meson decays. For these values, we use the benchmark values for $(W_u)_{1i}$ in Fig.~\ref{fig:Wu1i}, and $g_{B-L}$ which realizes BR$(K_L \to \pi^0 Z') = 2.1 \times 10^{-9}$ (the central value of the KOTO result) is used as the reference value.}
\label{tab:Bresult}
\end{table}
%%%%%%%%%%%%%%%%%%%%%%%%%%%%%%%%%%%%%%%%

Remarkably, the branching ratios related to $b \to s$ transition are about four orders of magnitude larger than those related to $b \to d$ transition. This feature is one of our interesting predictions in $B$ meson decays. Unfortunately, the current bounds for each decay mode are $\mathcal{O}(10^{-5})$, and therefore, the $b \to s$ transition is strongly constrained. In order to satisfy these constraints, $g_{B-L}$ needs to be about 30 times smaller than the current chosen value, $g_{B-L} = \mathcal{O}(10^{-3})$. In this case, the explanation of both $(g-2)_{\mu}$ and KOTO anomalies fails. However, the cancellation in the Scenario II does not completely pine down $(W_u)_{1i}$, which still leaves sufficient degrees of freedom to reduce $| g_{sbZ'}^{\rm eff}|$ by about one order, saving the Scenario II. We leave this issue to a future work. Note that in the Scenario I, the constraints of rare $B$ meson decays are satisfied since the required value of $g_{B-L}$ for the explanation of KOTO result is $\mathcal{O}(10^{-5})$. 

It is notable that the Belle II experiment aims to search the decay mode for $B \to K + E_{miss}$. 
The reported sensitivity on the branching ratio is about 10\% with 50ab$^{-1}$~\cite{Kou:2018nap}.

\section{Conclusions and discussions}
\label{sec:conclusion}

In this paper, we focus on the model in Ref.~\cite{Kang:2019vng}, in particular, the case where $Z'$ couples to the up-type quark flavor-dependently is considered. The model originally was designed to explain the $(g-2)_\mu$ anomaly via a muonic force carrier $Z'$. Although tree-level FCNCs in the down-type quark sector are forbidden by gauge symmetry, loop-level FCNCs are caused by the $W$ boson exchange but not taken into account in our previous study. The different point from the SM case is that flavor violating $Z'$ couplings exist, and therefore, the CKM suppression becomes mild. We calculated related loop diagrams and obtained the effective flavor-violating coupling for $s \to d$ transition, $g_{dsZ'}^{\rm eff}$. Then, by considering the $g_{dsZ'}^{\rm eff}$ contribution, we discuss its implications to the model, especially the possibility to explain the KOTO result, and the strong constraint on the viable parameter space for $(g-2)_\mu$. 

Because of this mild CKM suppression, the branching ratio for $K \to \pi Z'$ can be easily enhanced. 
For the generic $Z'$ mass, we found that the KOTO result can be explained with $g_{B-L} = \mathcal{O}(10^{-5})$, however, $K^+ \to \pi^+ \nu \bar{\nu}$ constraint gives $g_{B-L} < 5.4 \times 10^{-6}$. 
Then we cannot explain the KOTO result, and moreover, such small gauge coupling fails to explain the $(g-2)_{\mu}$ anomaly. Nevertheless, there are some mass windows where $K^+ \to \pi^+ \nu \bar{\nu}$ constraint should not be applied due to the huge background of $K^+ \to \pi^+ \pi^0$. When $m_{Z'} = 125 \sim 130$ MeV and $140 \sim 150$ MeV with appropriate structure of $W_u$, the mixings among the right-handed up-type quarks, the KOTO result can be explained with $g_{B-L} = \mathcal{O}(10^{-3})$ which is needed for the explanation of  $(g-2)_{\mu}$ anomaly. 
Especially, we found that the size of $(W_u)_{13}$ is very important for the explanation of the KOTO anomaly. 
Note that all possibilities of the explanation with $g_{B-L} \simeq \mathcal{O}(10^{-3}) \mathchar`- \mathcal{O}(10^{-5})$ can be tested by the COHERENT experiment by using non-standard neutrino interactions and the NA64 experiment by using muon beam.

However, such structure of $W_u$ and $g_{B-L} = \mathcal{O}(10^{-3})$ lead to large branching ratio for $B$ meson decays with $b \to s$ transition caused by corresponding effective coupling, $g_{sbZ'}^{\rm eff}$.
Then the structure of $W_u$ and/or size of $g_{B-L}$ are constrained. 
In other words, there is an explicit correlation between $g_{dsZ'}^{\rm eff}$ and $g_{sbZ'}^{\rm eff}$, through $W_u$ and $g_{B-L}$. Therefore, the $(g-2)_{\mu}$ and KOTO anomalies may be explained by global analysis, without conflicting with any constraints from FCNCs of down-type quark sector. 
This study will be done in the near future. 

In summary, contrary to the original intention, the $(B-L)_{23}$ gauge boson is no longer an attractive solution to the $(g-2)_\mu$ puzzle owing to the down-type quark FCNCs, but it is a natural candidate to account for the new KOTO anomaly. Moreover, the $(g-2)_\mu$ puzzle may be resolved in the sector to realize the correct neutrino mixings, and we will investigate this possibility in a future publication.

\section*{Acknowledgements}

This work is supported in part by the National Science Foundation of China (11775086).

\appendix

\section{A check}

In order to check the above loop calculation, it is useful to compare with one of the previous works by M.~Pospelov \cite{Pospelov:2008zw}. 
They consider the chiral perturbation theory to calculate the branching ratio for $K^+ \to \pi^+ X$ with $X$ being new vector boson. 
Note that in their setup, the SM fermions couple to $X$ through the kinetic mixing, parameterized by $\kappa$. 
Then, the amplitude is
\begin{align}
\mathcal{M}_{K \to \pi X} = \frac{e \kappa m_X^2}{(4 \pi)^2 m_{K^+}^2} ( k + p )^{\mu} \epsilon^X_{\mu} W(m_X^2),
\end{align}
where $k$ and $p$ are the kaon and pion momentum, $\epsilon^X_{\mu}$ is the polarization vector of $X$. 
Here, $W^2(m_X^2) \simeq 10^{-12} ( 3 + 6 m_X^2 / m_{K^+}^2 )$, and the details of this function is discussed in Ref.~\cite{DAmbrosio:1998gur}. 
The branching ratio calculated by above amplitude is
\begin{align}
{\rm BR}(K^+ \to \pi^+ X) = \frac{\alpha \kappa^2}{4 (4 \pi)^4} \frac{m_X^2 W^2}{\Gamma_{K^+} m_{K^+}} \left[ \lambda \left( 1, \frac{m_{\pi^+}^2}{m_{K^+}^2}, \frac{m_X^2}{m_{K^+}^2} \right) \right]^{3/2}.
\end{align}
Compared with our branching ratio in Eq.~\eqref{eq:BRKp}, the relation between our $g_{dsZ'}^{\rm eff}$ and $\kappa$ is obtained as
\begin{align}
| g_{dsZ'}^{\rm eff} | = \frac{2 e}{(4 \pi)^2} \frac{m_X^2}{m_{K^+}^2} \frac{\sqrt{W^2(m_X^2)}}{f_+^{K^+ \pi^+} (m_X^2)} \cdot \kappa \approx 3 \times 10^{-10} \times \kappa \left( \frac{m_X}{100 \, {\rm MeV}} \right)^2,
\label{eq:gdskapparela}
\end{align}
where the last relation is valid for $m_X < 200$ MeV. 
Note that in order to compare with their calculation, $| g_{dsZ'}^{\rm eff} |$ in Eq.~\eqref{eq:gdskapparela} should be the sum of diagonal part of Eq.~\eqref{eq:gdseff}. 
By setting $(g_L^u)_{ii} = (g_R^u)_{ii} = g' \, (i = 1, 2 ,3)$ as like the dark photon model, we can obtain the following value: 
\begin{align}
| g_{dsZ'}^{\rm eff} | \simeq ( 6.8 \times 10^{-7} ) \times | g' |,
\end{align}
which results in the relation between $g'$ and $\kappa$ as
\begin{align}
| g' | \simeq 7.8 \times 10^{-4} \kappa
\label{eq:gpkappa}
\end{align}
when $m_X = m_{\pi^0}$. 

According to their paper, the branching ratio can be expressed by
\begin{align}
{\rm BR}(K^+ \to \pi^+ X) \simeq 8 \times 10^{-5} \times \kappa^2 \left( \frac{m_X}{100 \, {\rm MeV}} \right)^2,
\end{align}
and $\kappa \lesssim 0.02$ is needed to satisfy the constraint of $K^+ \to \pi^+ X$ with $m_X = m_{\pi^0}$. 
In our notation, this leads to $| g' | \lesssim 1.6 \times 10^{-5}$. 
This result seems to be inconsistent with the result in Eq.~\eqref{eq:Kpconst}. 
The reason is that the dominant contribution in the above calculation is not considered in Eq.~\eqref{eq:gpkappa}, namely, $(C_L)_{12}^{ds}$ in Eq.~\eqref{eq:CL}. 
Taking into account all part of Eq.~\eqref{eq:CL}, the relation Eq.~\eqref{eq:gpkappa} becomes $| g' | \simeq 6.2 \times 10^{-8} \kappa$. 
Moreover, $| g' | = \frac{g_{B-L}}{3} V_{ud}^{\ast} V_{us} \sim \mathcal{O}(0.1) \times g_{B-L}$, and therefore, the constraint on $g_{B-L}$ from the result of Ref.~\cite{Pospelov:2008zw} becomes 
\begin{align}
g_{B-L} < \mathcal{O}(10^{-8}).
\end{align}
This constraint is same order as in Eq.~\eqref{eq:Kpconst}.

\section{Numerical values for $b \to q$ transitions}
\label{app:btoqnumeval}

In this appendix, we show the numerical values for loop contributions to $B$ physics. 
We show the analytical expressions for $b \to q$ ($q = d, s$) transition and the definition of $C_{L,R}^{qb}$ in Eq.~\eqref{eq:Bloop}. 
Then, the numerical values of $C_{L,R}^{qb}$ can be calculated by setting $q^2$ appropriately. 
For $q^2 = m_{\pi^0}^2$ as reference value, $C_{L,R}^{qb}$ are obtained as
\begin{align}
C_L^{db} &= \begin{pmatrix}
- 1.8 \times 10^{-3} + 5.2 \times 10^{-3} i & - 6.3 \times 10^{-2} & - 1.9 \\
4.1 \times 10^{-4} - 1.2 \times 10^{-3} i & 1.5 \times 10^{-2} - 9.2 \times 10^{-6} i & 0.43 - 2.8 \times 10^{-4} i \\
- 4.2 \times 10^{-5} + 4.8 \times 10^{-5} i & - 6.9 \times 10^{-4} - 2.8 \times 10^{-4} i & - 1.8 \times 10^{-2} - 7.1 \times 10^{-3} i
\end{pmatrix}, \label{eq:CLdb} \\
C_R^{db} &= \begin{pmatrix}
8.4 \times 10^{-12} - 1.4 \times 10^{-11} i & 6.3 \times 10^{-8} & 1.0 \times 10^{-5} \\
- 4.1 \times 10^{-10} + 1.2 \times 10^{-9} i & - 7.4 \times 10^{-6} + 4.7 \times 10^{-9} i & - 1.4 \times 10^{-3} + 8.6 \times 10^{-7} i \\
2.3 \times 10^{-10} - 2.6 \times 10^{-10} i & 2.1 \times 10^{-6} + 8.6 \times 10^{-7} i & 2.6 \times 10^{-3} + 1.0 \times 10^{-3} i
\end{pmatrix},
\label{eq:CRdb} \\
C_L^{sb} &= \begin{pmatrix}
- 4.1 \times 10^{-4} + 1.2 \times 10^{-3} i & - 1.5 \times 10^{-2} & - 0.44 \\
- 1.8 \times 10^{-3} + 5.2 \times 10^{-3} i & - 6.3 \times 10^{-2} - 2.1 \times 10^{-6} i & - 1.9 - 6.4 \times 10^{-5} i \\
9.1 \times 10^{-5} - 2.8 \times 10^{-4} i & 3.4 \times 10^{-3} - 6.4 \times 10^{-5} i & 8.7 \times 10^{-2} - 1.6 \times 10^{-3} i
\end{pmatrix}, \label{eq:CLsb} \\
C_R^{sb} &= \begin{pmatrix}
1.9 \times 10^{-12} - 3.1 \times 10^{-12} i & 1.4 \times 10^{-8} & 2.4 \times 10^{-6} \\
1.8 \times 10^{-9} - 5.1 \times 10^{-9} i & 3.2 \times 10^{-5} + 1.1 \times 10^{-9} i & 5.9 \times 10^{-3} + 2.0 \times 10^{-7} i \\
- 4.9 \times 10^{-10} + 1.5 \times 10^{-9} i & - 1.1 \times 10^{-5} + 2.0 \times 10^{-7} i & - 1.3 \times 10^{-2} + 2.4 \times 10^{-4} i
\end{pmatrix}.
\label{eq:CRsb}
\end{align}
Similar to $C_L^{ds}$, there is no CKM suppression for $(C_L^{db})_{13}$ and $(C_L^{sb})_{23}$, and therefore, these elements will be dominant contributions to related $B$ meson decays.

\end{document}